\documentclass[twocolumn,aps,prl,superscriptaddress]{revtex4-2}

\usepackage{amsmath}
\usepackage{graphicx}
\usepackage{MnSymbol}

\begin{document}

\title{Side-leakage of facemask}

\author{B.N.J. Persson}
\affiliation{PGI-1, FZ J\"ulich, Germany, EU}
\affiliation{www.MultiscaleConsulting.com}

\begin{abstract}
Face masks are used to trap particles (or fluid drops) in a porous material (filter) in order to avoid or reduce
the transfer of particles between the human lungs (or mouth and nose) and the external environment. 
The air exchange between the lungs and the environment is assumed to occur through the
facemask filter. However, if the resistance to air flow through the filter is high some air 
(and accompanied particles) will leak through the filter-skin interface.
In this paper I will present a model study of the side-leakage problem.
\end{abstract}

\maketitle

\setcounter{page}{1}
\pagenumbering{arabic}

%\pagestyle{empty}

%%%%%%%%%%%%%% main text %%%%%%%%%%%%%%%%
%\begin{multicols}{2}

%%%%%%%%%%%%%% main text %%%%%%%%%%%%%%%%

\vskip 0.3cm
{\bf 1 Introduction}

Face masks are used to trap particles in a porous material (filter) in order to avoid or reduce
the transfer of particles between the human lungs and the external environment\cite{review}. The 
filter usually consist of a sheet of randomly arranged fibers made from a 
polymer, e.g. polyethylene. The effectiveness of the filter will increase with increasing
thickness of the filter and with decreasing size of the open channels through the filter. However,
increasing the effectiveness of the filter will in general increase the resistance to the air flow
through the filter, which may result in uncomfortable breathing experience, or side leakage of air between the skin and the 
filter surface. Thus, a recent report demonstrates the potential risk of increased face-to-mask seal leakage when N95
filtering facepiece respirators (N95 FFR) are covered by surgical, cloth, or medical masks\cite{side}. 
In this paper I will present a more detailed model study of the side-leakage problem.

\begin{figure}[tbp]
\includegraphics[width=0.47\textwidth,angle=0]{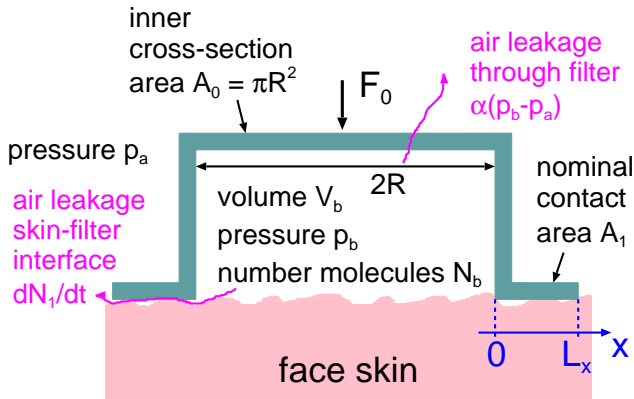}
\caption{
Model used for studying the air flow for facemask. The air can 
flow through the filter at a number flow rate $\alpha' (p_{\rm b} - p_{\rm a})$
or at the interface between the skin and the facemask at a number flow rate $dN_1/dt=\dot N_1$.
The number of air molecules $N_{\rm b}$ and the air pressure $p_{\rm b}$ in the volume $V_{\rm b}$
between the facemask and the face varies in time due to the breathing act.
}
\label{facepic.eps}
\end{figure}

\vskip 0.3cm
{\bf 2 Theory}

We consider the simplest (idealized) case where the facemask makes contact with the skin over a circular
annuls (radius $R$) of width $L_x$ in the radial (air leakage) direction and $L_y=2\pi R$ in the orthogonal angular direction
(see Fig. \ref{facepic.eps}). We assume the nominal contact pressure is constant in the 
nominal contact area $A_1 = L_x L_y$. 
The air volume between the face and the facemask is denoted by $V_{\rm b}$ and is assumed to be constant.
The facemask is pushed against the skin
by a force $F_0$ given by the extension of the rubber band, which is used to attach the facemask to the head,
and with the force $A_0 (p_{\rm b}-p_{\rm a})$ due to the air pressure difference between inside and outside the mask. 
The spring contact pressure $p_{\rm s} = F_0/A_1$. The nominal contact pressure in the area $A_1$ is
$$p=p_{\rm s}-\beta (p_{\rm b}-p_{\rm a}), \eqno(1)$$
where
$$\beta=A_0/A_1+\beta', $$
where $p_{\rm b}$ is the air pressure inside the facemask and $p_{\rm a}$ the air pressure outside,
which is assumed to be constant and equal to $1$ atm ($1 \ {\rm bar}$). 
The factor $\beta'$ is a number between $0.5$ and $1$ which depends on how the air pressure
change from $p_{\rm b}$ at $x=0$ to $p_{\rm a}$ at $x=L_x$ (see Fig. \ref{facepic.eps}).
Since $A_0/A_1 = \pi R^2/(2 \pi R L_x) = R/2L_x$ we have $\beta = R/2L_x + \beta'$. For N95 masks
the contact width $L_x$ is relative small so that $R/2L_x >> 1$ and $\beta'$ is not so important. 

The number of air molecules $N_b(t)$ inside the facemask satisfies
$$\dot N_b = \dot N(t) - \alpha' (p_{\rm b}-p_{\rm a}) - \dot N_1\eqno(2)$$
where $\dot N$ is the number of air molecules entering the volume $V_b$ from the lungs,
and $\alpha' (p_{\rm b}-p_a)$ the number of air molecules leaking though the facemask filter and
$\dot N_1$ the number of air molecules leaking between the skin and the facemask, which depends
on the nominal contact pressure $p$ and the pressure difference $\Delta p = p_{\rm b}-p_{\rm a}$ 
between inside and outside. 
We will assume the ideal gas law so that
$$p_{\rm b} V_{\rm b} = N_{\rm b} k_{\rm B} T\eqno(3)$$
The leakrate $\dot N_1$ is given approximately by\cite{tobe,Paolo} 
$$\dot N_1 = {1  \over 24} {L_y\over L_x} {(p_{\rm b}^2 -p_{\rm a}^2 )\over k_{\rm B}T} 
{u_{\rm c}^3  \over \eta }\eqno(4)$$
Here $k_{\rm B}T$ is the thermal energy ($k_{\rm B}$ is the Boltzmann constant and $T$ the absolute temperature), and
$u_{\rm c}$ is an effective surface separation which we determine using the Persson contact mechanics theory\cite{Yang,Alm,uc1}
and the Bruggeman effective medium theory as described elsewhere\cite{Dapp,seal1,seal2,seal3}.

The gas viscosity
$$\eta = {1\over 3} mn\bar v \lambda$$
where $n$ is the gas number density and $\lambda$ the mean free path due to collisions between gas molecules.
Note that $\lambda \sim 1/n$ so the viscosity $\eta$ is independent of the gas number density.
Equations (1)-(4) are 4 equations for the 4 unknown quantities, $p$, $p_{\rm b}$, $N_{\rm b}$ and $N_1$.

We denote the resistance to air flow through the facemask filter by $1/\alpha$, 
where the air flow conductance $\alpha$ is defined by
$${d V \over dt}= \alpha (p_{\rm b}-p_{\rm a})$$
where $\dot V = d V/dt$ is the volume of air of atmospheric pressure passing through the facemask filter
per unit time given the pressure difference $\Delta p = p_{\rm b}-p_{\rm a}$ between inside and outside 
the facemask. Using the ideal gas law we have
$$p_{\rm a} \dot V = \dot N k_{\rm B} T = \alpha' (p_{\rm b}-p_{\rm a}) k_{\rm B} T$$
so that $\alpha = \alpha' k_{\rm B} T/p_{\rm a}$.

\begin{figure}[tbp]
\includegraphics[width=0.47\textwidth,angle=0]{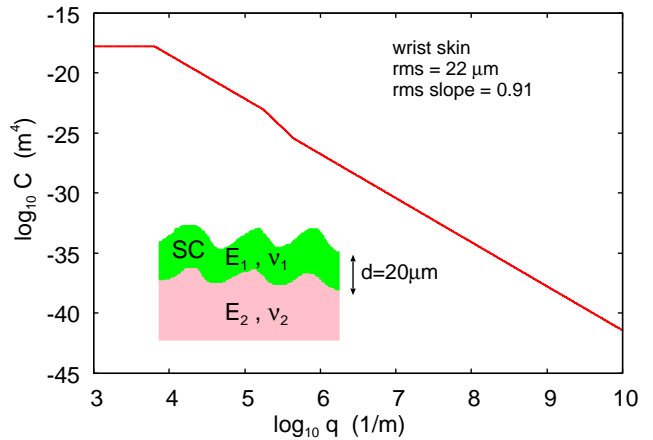}
\caption{
The surface roughness power spectrum as a function of the wavenumber (log-log scale)
obtained from optical and AFM measurements of the surface topography of 
the wrist skin of a 49 years old man (from Ref. \cite{Gorb1,Gorb2}). The surface has the rms roughness 
amplitude $22 \ {\rm \mu m}$ and the rms slope 0.91. The inset shows the skin model used in the
contact mechanics model calculations. The Young's modulus and Poisson ratio of the top layer
of the skin (stratum corneum, of thickness $d=20 \ {\rm \mu m}$) are $E_1=1 \ {\rm GPa}$ and
$\nu_1=0.5$, while the material below the top layer has $E_2= 20 \ {\rm kPa}$ and
$\nu_2=0.5$ (see \cite{Skin} for information about the human skin elastic properties).
}
\label{1logq.2logC.skin.eps}
\end{figure}

\begin{figure}[tbp]
\includegraphics[width=0.47\textwidth,angle=0]{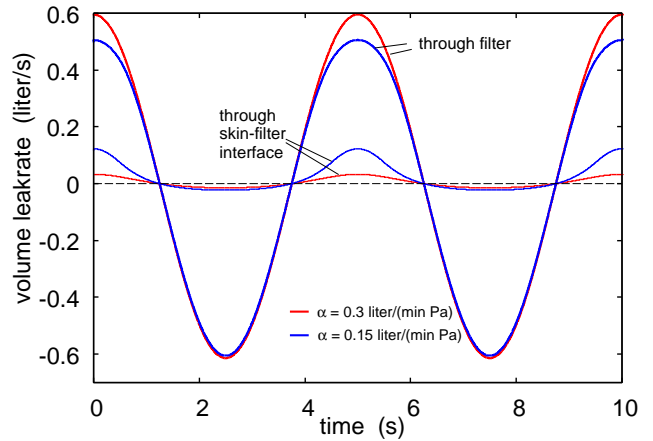}
\caption{
The air (of atmospheric pressure) volume leak rate in ${\rm liter/s}$ as a function of time
during breathing at the period of $T=5 \ {\rm sec}$ and the volume $V_0 = 0.5 \ {\rm liter}$:
$V(t)=V_0 {\rm sin}(\omega t)$ with $\omega = 2 \pi /T$. The red and blue curves are for the
filter air flow conductance $\alpha = 0.3$ and $0.15 \ {\rm liter/(min\cdot Pa)}$. The thick lines 
is the leakage through the filter and the thinner lines at the skin-filter interface.
}
\label{1time.VolumeLeakRate.eps}
\end{figure}

\begin{figure}[tbp]
\includegraphics[width=0.47\textwidth,angle=0]{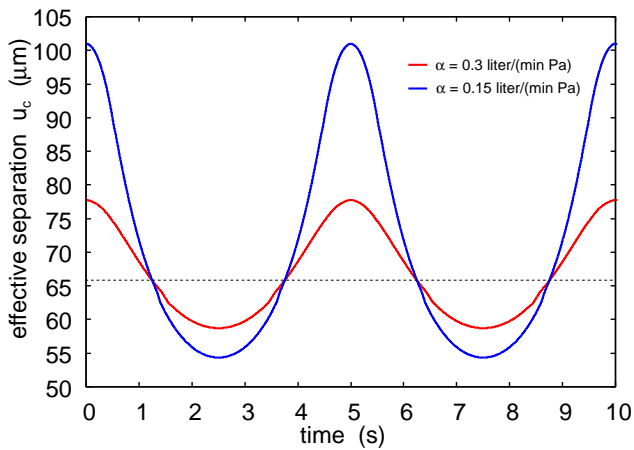}
\caption{
The effective surface separation as a function of time. 
}
\label{1time.2uc.eps}
\end{figure}

\begin{figure}[tbp]
\includegraphics[width=0.47\textwidth,angle=0]{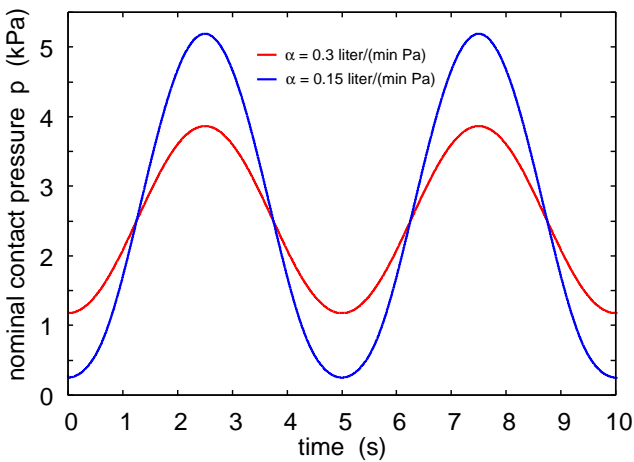}
\caption{
The nominal contact pressure in the filter-skin nominal contact area,
as a function of time.
}
\label{1time.2pcontact.kPa.eps}
\end{figure}

\begin{figure}[tbp]
\includegraphics[width=0.47\textwidth,angle=0]{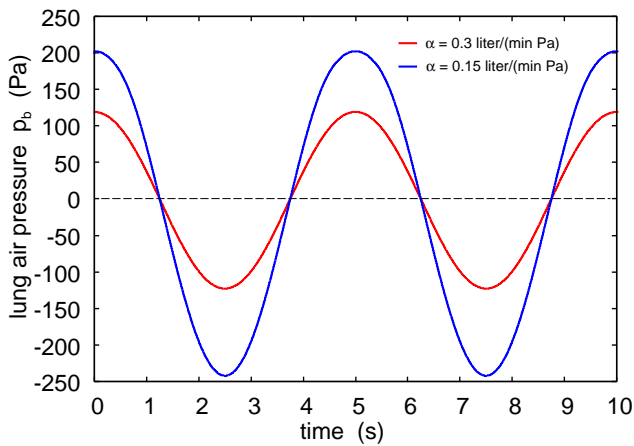}
\caption{
The air pressure (relative to the atmospheric pressure) in the lungs
as a function of time.
}
\label{1time.2pb.eps}
\end{figure}

\begin{figure}[tbp]
\includegraphics[width=0.47\textwidth,angle=0]{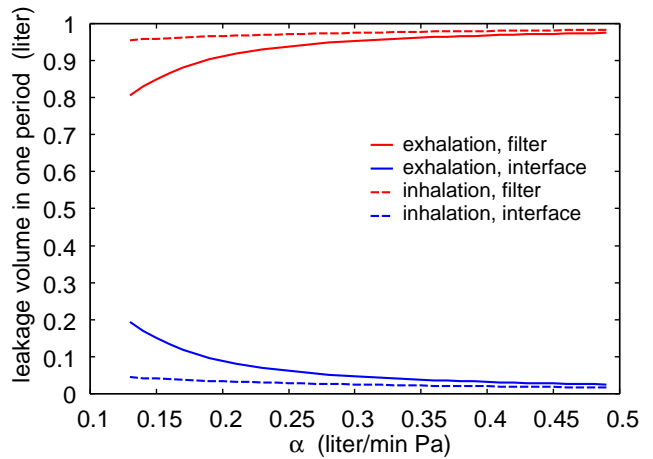}
\caption{
The volume of air passing through the filter (red lines) and through the
skin-filter interface (blue lines) as a function of the filter air flow conductance.
The solid lines is during exhalation and the dashed lines during inhalation.
}
\label{1alfa.2leakageVOLUME.eps}
\end{figure}

\vskip 0.3cm
{\bf 3 Numerical results}

We will assume that air molecules is injected and removed from the volume $V_{\rm b}$ by the breathing action
in a periodic way so that
$$N (t) = N_0 {\rm sin}(\omega_0 t)$$
corresponding to a volume of air (of atmospheric pressure) $V(t) = k_{\rm B}T N(t)/p_{\rm a}$,
$$V (t) = V_0 {\rm sin}(\omega_0 t)$$
where $\omega_0 = 2 \pi /T$ where $T$ is the period of breathing.
We assume that at time $t=0$, $p_{\rm b} = p_{\rm a}$ and hence from (4), $\dot N_1 = 0$ for $t=0$.

Ref. \cite{compare} present the air flow conductance of several types of facemasks.
For the US N95 facemask for the air flow $\dot V = 85 \ {\rm liter/min}$ 
the pressure difference $p_{\rm b}-p_{\rm a}$ should be smaller than  $343 \ {\rm Pa}$ 
during inhalation and $245 \ {\rm Pa}$ during exhalation (the two pressure drops may differ because of side leakage).
If the leakage would be entirely through the facemask these two cases correspond to
$\alpha \approx 0.25$ and $\alpha \approx 0.35 \ {\rm liter /min \cdot Pa}$.
For the European FFP2 facemask slightly larger (minimum) flow conductance are required.
Below we show results for $\alpha = 0.3$ and $0.15  \ {\rm liter /min \cdot Pa}$, where the smaller value
may reflect a N95 facemask contaminated by particles which block air flow channels. 

In the numerical study we use $T=5 \ {\rm s}$ and $V_0 = 0.5 \ {\rm liter}$, corresponding to
an air volume $1 \ {\rm liter}$ oscillating between the lungs and the outside of the lungs. 
We use the spring force $F_0=3 \ {\rm N}$ as measured for a N95 facemask on my head.
The facemask is assumed to make (nominal) contact with the skin over a circular strip of with
$L_x=3 \ {\rm mm}$ in the (radial) air leakage direction and of length $L_y = 2 \pi R = 40 \ {\rm cm}$
is the orthogonal angular direction. The volume between the face and the facemask is assumed to be
constant in time and equal to $0.1 \ {\rm liter}$. In reality the volume will fluctuate in time due to the
oscillations in the air pressure $p_{\rm b}(t)$, but this effect is not very important.

To calculate the air side leakage it is necessary to know the surface roughness on the facemask and the skin in the nominal contact
region. Here we will use the surface roughness measured on the wrist skin in Ref. \cite{Gorb1,Gorb2}. We expect the roughness
on the skin on the face to be similar but we have not studied it. 
We note, however, that if the skin is covered by hair (beard, poorly shaved or unshaved) this could effectively strongly increase the
surface roughness and result in much larger leakage rates than predicted below\cite{beard}.
In this section we will assume that the facemask surface
has no surface roughness. Including the surface roughness on the facemask surface will increase the side leakage.

Fig. \ref{1logq.2logC.skin.eps} shows
the surface roughness power spectrum as a function of the wavenumber (log-log scale)
obtained from optical and AFM measurements of the surface topography of 
the wrist skin of a 49 years old man (from Ref. \cite{Gorb1,Gorb2}). The surface has the root-mean-square (rms) roughness 
amplitude $22 \ {\rm \mu m}$ and the rms slope 0.91. The inset shows the skin model used in the
contact mechanics model calculations. The Young's modulus and Poisson ratio of the top layer
of the skin (stratum corneum, of thickness $d=20 \ {\rm \mu m}$) are $E_1=1 \ {\rm GPa}$ and
$\nu_1=0.5$, while the material below the top layer has $E_2= 20 \ {\rm kPa}$ and
$\nu_2=0.5$ (see \cite{Skin} for information about the human skin elastic properties).

Fig. \ref{1time.VolumeLeakRate.eps} shows 
the air (of atmospheric pressure) volume leak rate in ${\rm liter/s}$ as a function of time
during breathing. The red and blue curves are for the
filter air flow conductance $\alpha = 0.3$ and $0.15 \ {\rm liter/(min\cdot Pa)}$. 
The thick lines is the leakage through the filter and the thinner lines at the skin-filter interface.
Note that the side air leakage is larger during exhalation than during inhalation. This is due to the 
air pressure term $(p_{\rm b}-p_{\rm a})A_0$ which increases the force squeezing the facemask against the skin during
inhalation while it reduces the contact force during exhalation. For $\alpha = 0.3 \ {\rm liter/(min\cdot Pa)}$
for exhalation about $5\%$ of the air leaks between the skin and the facemask while during inhalation only
$2.6\%$ of the air side leaks. For $\alpha = 0.15 \ {\rm liter/(min\cdot Pa)}$ the corresponding numbers are
$18\%$ and $4\%$.

Fig. \ref{1time.2uc.eps} shows the effective surface separation $u_{\rm c}$ as a function of time.
The dotted line indicate the effective surface separation when there is no pressure force, i.e., when 
$p_{\rm a}=p_{\rm b}$. As expected, during exhalation the surface separation increases while it decreases
during inhalation. Similarly, the nominal contact pressure in the filter-skin nominal contact area 
decreases during exhalation and increases during inhalation as shown in
Fig. \ref{1time.2pcontact.kPa.eps}.
Fig. \ref{1time.2pb.eps} shows the air pressure (relative to the atmospheric pressure) 
in the facemask volume $V_{\rm b}$ as a function of time.

Finally, in Fig \ref{1alfa.2leakageVOLUME.eps}
we show the volume of air passing through the filter (red lines) and through the
skin-filter interface (blue lines) as a function of the filter air flow conductance.
The solid lines is during exhalation and the dashed lines during inhalation. For small 
air filter conductance the side leakage is large during exhalation while during
inhalation there is a much smaller side leakage.

\begin{figure}[tbp]
\includegraphics[width=0.4\textwidth,angle=0]{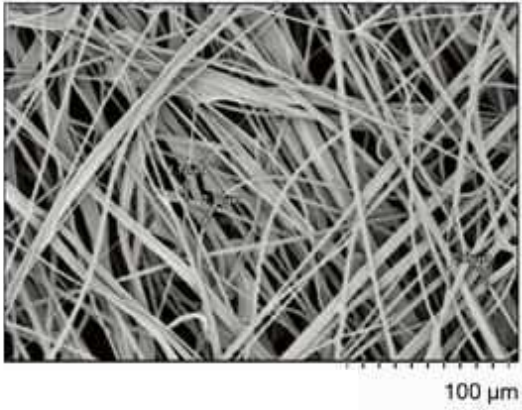}
\caption{
Scanning electron micrograph image of filter layer (non-woven polypropylene, melt blown).
Adapted from Ref. \cite{SEM}.
}
\label{fiber.ps}
\end{figure}

\begin{figure}[tbp]
\includegraphics[width=0.47\textwidth,angle=0]{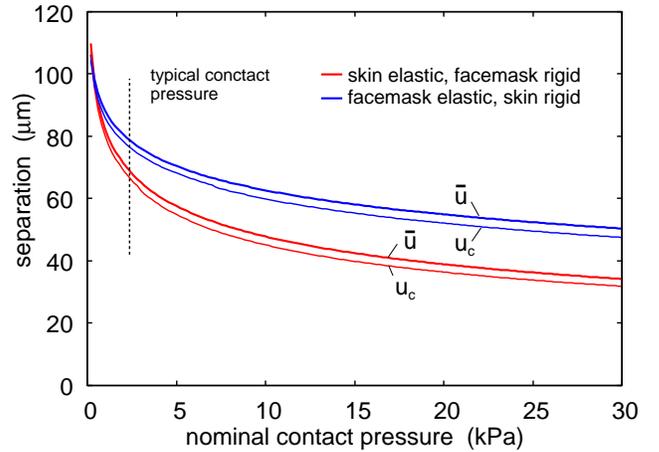}
\caption{
The average interfacial separation $\bar u$, and the separation $u_{\rm c}$ which determines the
air leakage [see (4)], as a function of the nominal contact pressure. The red lines are the results
assuming the facemask is rigid and the skin 
elastic (with layered elastic properties, see Fig. \ref{1logq.2logC.skin.eps}),
and the blue lines is the result assuming
the facemask elastic (with thickness $1 \ {\rm mm}$ and with $E=27.5 \ {\rm MPa}$ and $\nu=0.37$) and the skin rigid. 
Only the surface roughness of the skin (with the power spectrum shown in Fig.  \ref{1logq.2logC.skin.eps})
is included in the calculations.
}
\label{1pressure.2uc.and.uav.two.cases.eps}
\end{figure}

\begin{figure}[tbp]
\includegraphics[width=0.47\textwidth,angle=0]{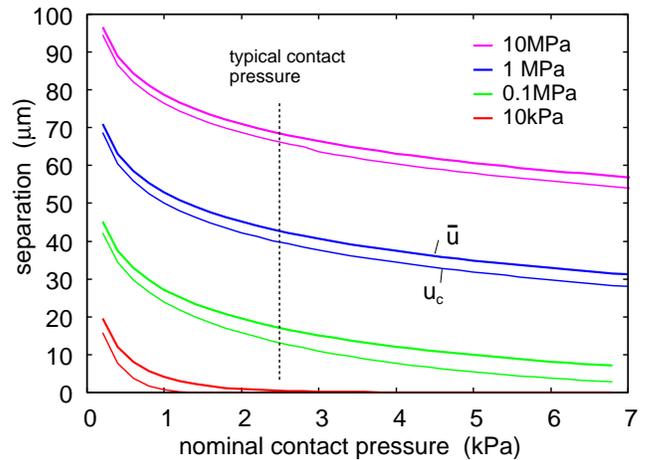}
\caption{
The average interfacial separation $\bar u$, and the separation $u_{\rm c}$ which determines the
air leakage [see (4)], as a function of the nominal contact pressure. Results are shown for the elastic
modulus $E=10$, $1$, $0.1$ and $0.01 \ {\rm MPa}$. I all cases the Poisson ratio $\nu=0.5$. 
Only the surface roughness of the skin (with the power spectrum shown in Fig.  \ref{1logq.2logC.skin.eps})
is included in the calculations.
}
\label{1pressure.2baru.and.uc.10MPa.1MPa.0.1MPa.0.01MPa.eps}
\end{figure}

\begin{figure}[tbp]
\includegraphics[width=0.3\textwidth,angle=0]{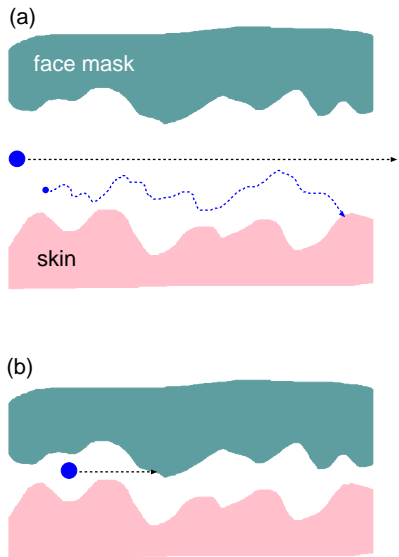}
\caption{
A large particle or droplet, due to its large inertia, 
will not be able to respond to the rapidly fluctuating (due to the surface roughness)
air flow current and will hence move through the skin-filter nominal contact
region on a nearly strait line. (a) If the average spacing between the surfaces in the air flow channels
are much larger than the surface roughness amplitude then big particles may pass through the 
contact without collisions with the walls. (b) If the average spacing is of order, or smaller than, the
surface roughness amplitude the big particle is likely to hit into the solid walls. A very small
particle will perform Brownian motion in addition to drifting with the air flow. In this case if the Brownian
motion amplitude is big enough the particle may hit into a wall even if the average wall separation is large as in
(a). 
}
\label{Movement.eps}
\end{figure}

\begin{figure}[tbp]
\includegraphics[width=0.3\textwidth,angle=0]{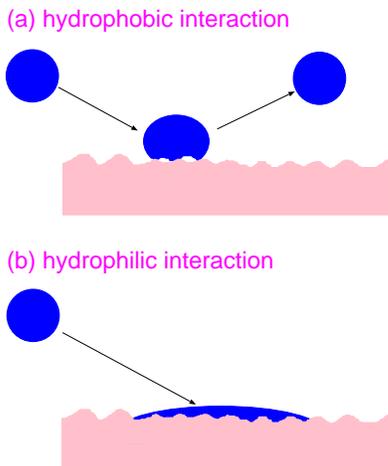}
\caption{
(a) If the fluid-solid interaction is hydrophobic a liquid droplet hitting a solid may
bounce off without transfer of fluid to the solid wall\cite{De,EPL}. (b) If the interaction is hydrophilic
the droplet may be adsorbed on the solid wall. 
}
\label{HydroPhobic.eps}
\end{figure}

\begin{figure}[tbp]
\includegraphics[width=0.35\textwidth,angle=0]{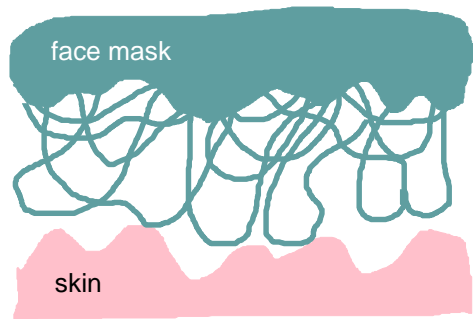}
\caption{
The N95 mask has a thin layer of fibers pointing away from the facemask surface
which may help to trap particles in the side leakage air flow channel.
}
\label{Fiber1.eps}
\end{figure}

\vskip 0.3cm
{\bf 4 Discussion}

The study above is very idealized as we have assumed a uniform contact pressure in the 
facemask-skin nominal contact area. We have also neglected the surface roughness on the 
facemask and treated the facemask as rigid when calculating the effective surface separation 
$u_{\rm c}$ which determined the leakrate. 

The N95 facemask is built from non-woven polymer
fibers, e.g., polypropylene, polyethylene or polyesters (see Fig. \ref{fiber.ps})\cite{review}. 
(Recently it has been suggested to instead use polymer films with a periodic distribution of closely spaced 
nanoholes as facemask filters\cite{nano}.)
The polymers have a large elastic modulus (typically several GPa) but the fiber mat
is macroscopically relative soft with an effective modulus in tension typically in the range $10-100 \ {\rm MPa}$.
The fiber mat can be treated as an homogeneous material only on length scales larger than both the fiber
width (or thickness), and the average distance between two nearby fiber segments which typically means distances of 
order $10 \ {\rm \mu m}$ or more. However, because of the low nominal contact pressure, 
the average and effective surface separation in the present case are 
determined by the most long wavelength surface roughness
components so to a good approximation we can treat the fiber mat as a 
homogeneous material. Another problem is that the fiber mat
cannot in general be treated as a isotropic elastic material but the effective modulus $E_z$ which determine 
the elongation (tension) normal to the film, which is important for the
contact mechanics, may differ from the elastic modulus $E_x=E_y$ in tension within the film plane. 
However, if the fibers bind to each other where they touch each other, and if they are closely spaced, 
we expect the effective modulus $E_z$ to be similar to the modulus in tension within the plane.
% (see Appendix A).
 
We now show that including the macroscopic elastic properties of the 
facemask material, by using the modulus obtained in tension, will not result in any drastic change in the 
results presented above. To illustrate this,
in Fig. \ref{1pressure.2uc.and.uav.two.cases.eps}
we show the average interfacial separation $\bar u$, and the separation $u_{\rm c}$ which determines the
air leakage [see (4)], as a function of the nominal contact pressure. The red lines are the results
assuming the facemask is rigid and the skin 
elastic (with layered elastic properties, see Fig. \ref{1logq.2logC.skin.eps}),
and the blue lines is the result assuming
the facemask elastic (with $E_x=27.5 \ {\rm MPa}$ and $\nu=0.37$, as measured for 
non-woven polypropylene in tension\cite{Lei,Schaff}) 
and the skin rigid. 
In the latter case we have assumed the facemask has the thickness $1 \ {\rm mm}$ but practically the same result is obtained
assuming infinite thickness.
Only the surface roughness of the skin (with the power spectrum shown in Fig.  \ref{1logq.2logC.skin.eps})
is included in the calculations. 

We will show below that in order to effective trap particles and droplets the surface separation in the nominal skin-facemask
contact region should be much smaller than shown in Fig. \ref{1pressure.2uc.and.uav.two.cases.eps}. 
This can be realized if the rim of the facemask is covered by a strip of a soft material, e.g., weakly crosslinked Polydimethylsiloxan (PDMS).
To illustrate what elastic modulus is necessary, in Fig. \ref{1pressure.2baru.and.uc.10MPa.1MPa.0.1MPa.0.01MPa.eps}
we show the average interfacial separation $\bar u$, and the separation $u_{\rm c}$ which determines the
air leakage [see (4)], as a function of the nominal contact pressure. Results are shown for the elastic
modulus $E=10$, $1$, $0.1$ and $0.01 \ {\rm MPa}$. I all cases the Poisson ratio $\nu=0.5$. 
Only the surface roughness of the skin (with the power spectrum shown in Fig.  \ref{1logq.2logC.skin.eps})
is included in the calculations. Note that for $E<10 \ {\rm kPa}$ the contact area percolate at a pressure
below the average pressure in the skin-facemask contact region, so for such soft
material no leakage of air or particles would be possible. In the calculations we have neglected adhesion which is
important for the leakage only for very soft materials with $E<0.1 \ {\rm MPa}$ (see Appendix A).
Thus including adhesion for the $E= 10 \ {\rm kPa}$ case in Fig. \ref{1pressure.2baru.and.uc.10MPa.1MPa.0.1MPa.0.01MPa.eps} 
would reduce reduce $\bar u$ and $u_{\rm c}$. 

A weakly crosslinked PDMS would be very sticky and perhaps uncomfortable to use and 
may get contaminated by dust particles.
Another possibility would be to use an elastically soft hydrogel\cite{Gong,Sawyer} strip
at the edge of the facemask, with a thickness of order a few mm. A proper chosen hydrogel may exhibit no or negligible
adhesion to the human skin\cite{Gong1}. However, hydrogels will dehydrate and becomes rigid in air but perhaps the moisture
in the air from the lungs is enough to keep it hydrated.

So far we have only considered the air flow problem. The question is now if particles in the air
are able to follow the air flow into or out of the facemask volume $V_{\rm b}$. For the flow through the 
facemask filter this problem has been studied in detail both theoretically and experimentally. 
Several different mechanisms have been proposal which result in trapping of particles in the filter:

(A) Inertial Impacting: Aerosol or dust particles typically $1 \ {\rm \mu m}$ or larger in size with enough 
inertia to prevent them from flowing around the fibers in the filtration 
layers slam into the mask material where they may adhere (but see below) and get filtered.

(B) Diffusion: Particles smaller than $1 \ {\rm \mu m}$, usually $0.1 \ {\rm \mu m}$ and smaller 
that are not subject to inertia undergo diffusion and get stuck to fibrous layers of the filter 
while undergoing Brownian motion around the tortuous porous matrix of the filter fiber.

(C) Electrostatic attraction: This mechanism employs electrocharged polymer or resin fibers 
that attract both large and small oppositely charged particles, or neutral particles via polarization (induced dipole) effects, 
and trap them. This effect depends on the distribution of positive and negative charges on the polymer surfaces
(the total charge is likely to vanish so there must be an equal number of positive and 
negative charges on the polymer fibers)\cite{electroadhesion}.

The critical or equivalent pore diameter (see Appendix B and Ref. \cite{Lind}) in currently available N95 masks 
are around $300 \ {\rm nm}$ in size, while the SARS-CoV-2 virus is significantly 
smaller at 65 to $125 \ {\rm nm}$. However, the virus always travels attached to larger particles that are consistently snared by the filter. 
Thus, the virus are usually attaches to water droplets or aerosols (i.e. really small droplets) that are generated by breathing, 
talking, coughing, etc. These consist of water, mucus protein and other biological material and are all 
of order or larger than $1 \ {\rm \mu m}$. And even if the particles were smaller than the N95 filter size, 
the erratic Brownian motion of particles that size and the 
electrostatic attraction generated by the mask means they would be consistently caught as well.

The fibers in facemasks are usually made from a hydrophobic polymer in order to avoid the facemask absorb
moisture from the air from the lungs. However, experiments have shown that a water droplet hitting a hydrophobic 
surface may bounce off which would result in a reduced trapping of fluid (aerosols) 
droplets (see Fig. \ref{HydroPhobic.eps})\cite{De,EPL}. 
Thus an interesting problem is to find out which water contact angle
is optimal in order to avoid (or reduce) water absorption from the humid air from the lungs 
but still allow water droplets to get stuck to the fibers during impact from the air.
Once stuck to a fiber it is also important what happens to a respiratory 
droplet (e.g. evaporation of water) as this may effect the time
period a trapped virus (or bacteria) is intact or alive\cite{droplet}. This too will depend on the 
chemical nature (and the surface topography) of the fiber material.

The N95 and FFP2 facemasks have thin layers of fibers pointing away from the facemask surfaces
which may help to trap particles in the side leakage air flow channel (see Fig. \ref{Fiber1.eps}).

Experiments have shown that N95 masks are actually best for particles either larger or smaller than the $300 \ {\rm nm}$ threshold. Thus
N95 masks actually have that name because they are 95\% efficient at stopping particles in their least efficient 
particle size range in this case those around $0.3 \ {\rm \mu m}$.
Thus, particles smaller than $\sim 1 \ {\rm \mu m}$ perform erratic, zig-zagging Brownian motion with large enough amplitude
to hit into a fiber, which greatly increases the chance they will be snared by the mask fibers.

The trapping mechanisms (A)-(C) are also relevant for trapping if particles in the air stream between the skin and the facemask.
However, the effective separation between the skin and the facemask is much larger then the $0.3 \ {\rm \mu m}$ pore size
in the facemask filter. Thus Fig. \ref{1time.2uc.eps} shown that $u_{\rm c}$ is typically  between $50-100  \ {\rm \mu m}$ which is 
several times bigger than the skin rms roughness amplitude (about $20 \ {\rm \mu m}$). Hence it is possible for micrometer
sized particles to pass through the skin-facemask contact region as indicated in Fig. \ref{Movement.eps}(a). In order for the
inertia effect trapping mechanism to be effective one would need the average surface separation to be of order the rms
surface roughness amplitude as indicated in Fig. \ref{Movement.eps}(b). Furthermore, the Brownian motion trapping mechanism 
(B) (see Fig. \ref{Movement.eps}(a)) may be ineffective to trap small particles. To see this note that during a time $t$ the mean square
displacement, due to Brownian motion of a spherical particle (radius $R$), in one direction is given by\cite{Ein}
$$\langle x^2 \rangle = {k_{\rm B}T t \over 3 \pi \eta R}\eqno(5)$$
The (average) air flow velocity $v$ in the skin-facemask interfacial region is given by
$\dot V =  v L_y \bar u$ where $\dot V$ is the volume rate of air leakage at the interface and $\bar u$ the average surface separation.
Using $\dot V =0.03 \ {\rm liter/s}$ and $\bar u = 70 \ {\rm \mu m}$ gives $v \approx 1 \ {\rm m/s}$.
We assume the Brownian particle drift with the air stream so the time in the interfacial region will be $t = L_x/v$
or $t \approx 3\times 10^{-3} \ {\rm s}$ where we have used $L_x=3 \ {\rm mm}$.
Using the air viscosity $\eta \approx 2\times 10^{-5} \ {\rm Pas}$ and
$\langle x^2 \rangle \approx \bar u^2$ we get
$$R \approx {k_{\rm B}T t \over 3 \pi \eta \bar u^2} \approx 10^{-11} \ {\rm m}$$
Thus, trapping of particles resulting from Brownian motion is negligible in 
the side leakage channel. This is different in the facemask filter where the open channels may on the average have a diameter
of order $1 \ {\rm \mu m}$ (and the most narrow constriction may be only $0.3 \ {\rm \mu m}$). 
This will enhance $R$ by a factor $(70/1)^2 \approx 5000$ which will make Brownian motion important
for particles smaller than $\sim 0.1 \ {\rm \mu m}$.

\vskip 0.3cm
{\bf 5 Summary and conclusion}

We have studied a very simple model for the side leakage of facemasks. We have assumed that the 
skin-facemask nominal contact pressure is the same everywhere in the nominal contact area, and neglected
the surface roughness of the facemask surface. The calculations indicate that under normal conditions
for the N95 or FFP2 facemasks a few \%  of the air may leak through the skin-facemask interface. The average separation between the
surfaces in the skin-facemask contact region is much larger than the effective pore size in the facemask
filter which allow suspended particles in the air to enter or leave the facemask volume $V_{\rm b}$
during inhalation and exhalation.

\vskip 0.5cm
{\bf Acknowledgments:}

I thank J.P. Gong, M. Scaraggi and E. Tosatti for useful comments on the manuscript.

%\vskip 0.3cm
%{\bf Appendix A: effective elastic modulus of non-woven fiber mat} 

%...

\begin{figure}[tbp]
\includegraphics[width=0.47\textwidth,angle=0]{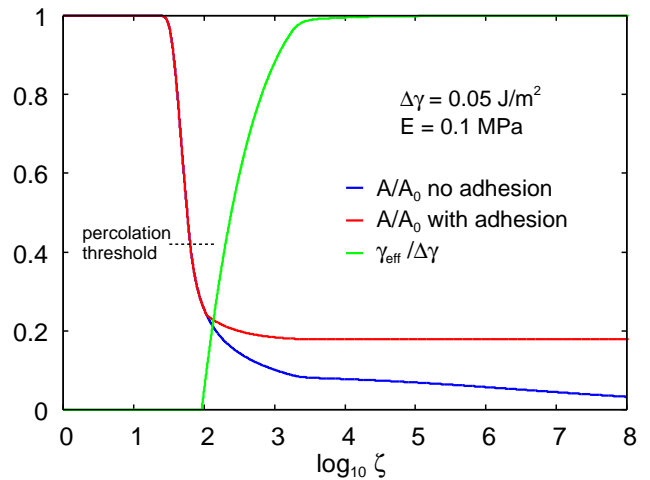}
\caption{
The relative contact area $A/A_0$ with (red line) and without (blue line) adhesion,
and the effective interfacial binding energy $\gamma_{\rm eff}$ (or work of adhesion) (green line)
as a function of the logarithm of the magnification. We have used the elastic modulus
$E=0.1 \ {\rm MPa}$ and the work of adhesion for smooth 
surfaces $\Delta \gamma = 0.05 \ {\rm J/m^2}$. Note that the contact area percolate
at a magnification where the adhesion does not manifest itself.
}
\label{1logz.2Area.and.gamma.for.E=0.1MPa.1.eps}
\end{figure}

\begin{figure}[tbp]
\includegraphics[width=0.47\textwidth,angle=0]{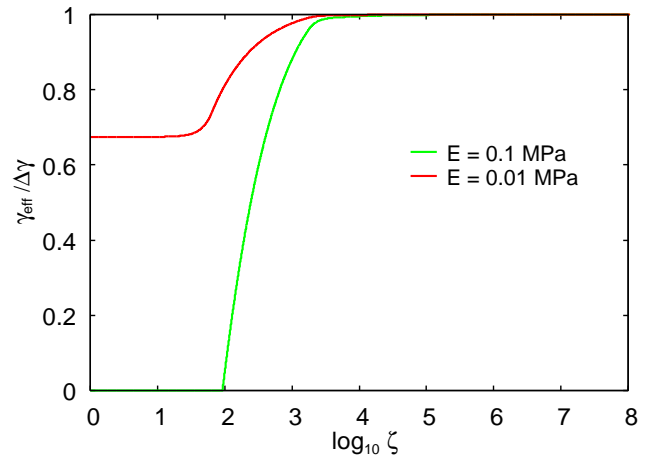}
\caption{
The effective interfacial binding energy $\gamma_{\rm eff}$ (or work of adhesion)
as a function of the logarithm of the magnification. We have used the elastic modulus
$E=0.1 \ {\rm MPa}$ (green line) and  $E=0.01 \ {\rm MPa}$ (red line), and the work of adhesion for smooth 
surfaces $\Delta \gamma = 0.05 \ {\rm J/m^2}$. Note that $\gamma_{\rm eff}$ vanish
for $\zeta = 1$ when $E=0.1 \ {\rm MPa}$ while it is nonzero when $E=0.01 \ {\rm MPa}$.
Thus only in the latter case will there be a finite pull-off 
force in the adiabatic (infinitely slowly) pull-off limit.
}
\label{1logz.2gamma.E=0.1MPa.E=0.01MPa.eps}
\end{figure}

\vskip 0.3cm
{\bf Appendix A: role of adhesion} 

In calculating the air leakrate we have used the effective medium approach combined
with the Persson contact mechanics theory for the probability distribution of surface separations.
The basic contact mechanics picture (critical junction theory) which can be used to estimate the leak-rate
of seals is as follows: Consider first a seal where the nominal contact area
is a square. The seal separate a high-pressure gas 
on one side from a low pressure gas on the other side, with the pressure drop $\Delta P$.
We consider the interface between the solids at increasing
magnification $\zeta$. At the magnification $\zeta$ only roughness components with wavenumber
$q<\zeta q_0$ can be observed, where $q_0$ is the smallest wavenumber.
At low magnification we observe no surface roughness and it appears
as if the contact is complete (see blue line in Fig. \ref{1logz.2Area.and.gamma.for.E=0.1MPa.1.eps}). 
Thus studying the interface only at this low
magnification we would be tempted to conclude that the leak-rate
vanishes. However, as we increase the magnification $\zeta$
we observe surface roughness and non-contact regions, so that
the contact area $A(\zeta)$ is smaller than the nominal contact area $A_0 = A(1)$. As we increase
the magnification further, we observe shorter wavelength roughness, and $A(\zeta)$ decreases
further. For randomly rough surfaces, as a function of increasing magnification, when
$A(\zeta)/A_0 \approx 0.42$ the non-contact area percolate\cite{Dapp}, and the first open channel is observed,
which allow fluid to flow from the high pressure side to the low pressure side.
The percolating channel has a most narrow constriction over which most of the pressure drop
$\Delta P$ occur. In the simplest picture one assume that the whole
pressure drop $\Delta P$ occur over this {\it critical constriction}, and if it is
approximated by a rectangular pore of height $u_{\rm c}$ much smaller than its width $w$
(as predicted by contact mechanics theory), the leak rate can be approximated by Eq. (4).
The height $u_{\rm c}$ of the critical constriction can
be obtained using the Persson contact mechanics theory
(see Ref. \cite{Yang,Alm,uc1}). 

When adhesion is included the average interfacial separation $\bar u$ and $u_{\rm c}$ decreases.
However, the influence of adhesion on these quantities is in the present case negligible when the
elastic modulus $E>0.1 \ {\rm MPa}$. This can be understood by studying how the contact are decreases
with the magnification. 
Fig. \ref{1logz.2Area.and.gamma.for.E=0.1MPa.1.eps}
shows the relative contact area $A/A_0$ with (red line) and without (blue line) adhesion,
and the effective interfacial binding energy $\gamma_{\rm eff}$ (or work of adhesion) (green line)
as a function of the logarithm of the magnification\cite{adhesion1,adhesion2}. We have used the elastic modulus
$E=0.1 \ {\rm MPa}$ and the work of adhesion for smooth 
surfaces $\Delta \gamma = 0.05 \ {\rm J/m^2}$ (as typical for adhesion involving PDMS). 
Note that the contact area percolate
at a magnification where the adhesion does not manifest itself.

Fig. \ref{1logz.2gamma.E=0.1MPa.E=0.01MPa.eps}
shows the effective interfacial binding energy $\gamma_{\rm eff}$ (or work of adhesion)
as a function of the logarithm of the magnification. We have used the elastic modulus
$E=0.1 \ {\rm MPa}$ (green line) and  $E=0.01 \ {\rm MPa}$ (red line), and the work of adhesion for smooth 
surfaces $\Delta \gamma = 0.05 \ {\rm J/m^2}$. Note that $\gamma_{\rm eff}$ vanish
for $\zeta = 1$ when $E=0.1 \ {\rm MPa}$ while it is nonzero for $\zeta = 1$ when $E=0.01 \ {\rm MPa}$.
Thus only in the latter case will there be a finite pull-off force 
in the adiabatic (infinitely slowly) pull-off limit.

We note finally that for some hydrogel--countersurface systems it is possible to avoid adhesion\cite{Gong1},
in particular at the low contact pressures of interest here. The effective repulsion in these systems
is an entropy effect related to squeezing out water from the interface resulting in a higher concentration
of the solvated ions in the water, which screen surface charges bound to the gel surface.

\vskip 0.3cm
{\bf Appendix B: effective pore size} 

Filters made from non-woven polymer fibers have complex irregular air flow channels and no 
well defined pore size. For these systems one speak about equivalent pore diameter which can be determined
using the bubble-point test as described in Ref. \cite{Lind}. 

Consider all percolating open channels through the
filter. For each channel there will be a most narrow constriction. The equivalent pore diameter is a length
characterizing the width of the biggest of all these most narrow constrictions\cite{Lind}. 
It can also be considered as the diameter of the critical
constriction as introduced in the theory of seals developed in 
Ref. \cite{seal1,seal2,seal3}. In this theory the system is studied
at increasing magnification or resolution. At the lowest magnification (say naked eye) the fiber mat appears 
as a homogeneous material and no leakage would be expected at this magnification. As we increase the magnification we
observe some ``big'' cavity regions but they do not percolate so even at this magnification one would not expect any leakage.
When we increase the magnification even more (say using an optical microscope)
one finally observe for the first time an open percolating path. The critical
constriction is the most narrow constriction along the first percolating path which can be observed 
with increasing magnification. The equivalent pore diameter is roughly 
the diameter of a circular hole with the same cross section area as that of the critical constriction.

\vskip 0.2cm 
{\bf Data Availability Statement:}
Data sharing is not applicable to this article as no new data 
were created or analyzed in this study.

\end{document}